\documentclass[aps, twocolumn, groupedaddress, superscriptaddress, floatfix, longbibliography]{revtex4-1}
\usepackage{amsmath}
\usepackage{amssymb}
\usepackage{graphicx}
\usepackage{float}
\usepackage[english]{babel}
\usepackage{dsfont}
\usepackage{epstopdf}
\usepackage{soul}
\usepackage{color}
\usepackage{braket}
\usepackage[normalem]{ulem}
\usepackage[colorlinks=true,
            linkcolor=blue,
            urlcolor=blue,
            citecolor=blue]{hyperref}

\definecolor{teal}{rgb}{0,0.44,0.5}
\definecolor{SMblue}{rgb}{1.0,0.2,1.0}%{0.0, 0.0, 0.0}%
\definecolor{SMblue2}{rgb}{0.0, 0.0, 0.0}%{0.0,0.2,1.0}%% 
\definecolor{TriClr}{rgb}{0,0,0}%{0,0,0} %{0.6,0.3,1.0}

\emergencystretch=\maxdimen
\newcommand{\dd}[3][]{\frac{\text{d}^{#1} #2}{\text{d} {#3}^{#1}}}

\newcommand{\bp}[1]{\left( #1 \right)}

\newcommand{\bs}[1]{\left[ #1 \right]}
\newcommand{\creop}[1]{\hat{a}^{\dagger}_{#1}}
\newcommand{\aniop}[1]{\hat{a}_{#1}}
\begin{document}

\title{Intensity Correlation Measurement to Simulate Two-body BICs\\ and Probe Nonlinear Discrete Breathers}

\author{Trideb Shit}
\affiliation{Department of Physics, Indian Institute of Science, Bangalore 560012, India}
\author{Rishav Hui}
\affiliation{Department of Physics, Indian Institute of Science, Bangalore 560012, India}
\author{Marco Di Liberto}
\affiliation{Dipartimento di Fisica e Astronomia ``G. Galilei" \& Padua Quantum Technologies Research Center, Universit\`a degli Studi di Padova, I-35131 Padova, Italy}
\affiliation{Istituto Nazionale di Fisica Nucleare (INFN), Sezione di Padova, I-35131 Padova, Italy}
\author{Diptiman Sen}
\affiliation{Centre for High Energy Physics, Indian Institute of Science, Bangalore 560012, India}
\author{Sebabrata Mukherjee}
\email{mukherjee@iisc.ac.in}
\affiliation{Department of Physics, Indian Institute of Science, Bangalore 560012, India}
%\date{\today}

\begin{abstract}
We study Hanbury Brown-Twiss spatial intensity correlations in femtosecond laser-fabricated photonic Su-Schrieffer-Heeger lattices using coherent input states with tunable phases. 
By mapping intensity correlations to the two-body quantum walk, we experimentally simulate edge bound states in the continuum (BICs) of two indistinguishable bosons. These two-body edge BICs show remarkable robustness in the presence of disorder. We then discuss how intensity correlations can capture the dynamics of two identical fermions for which BICs do not exist on the same edge of the lattice. The localization of intensity correlation, observed in the linear regime, persists at weak nonlinearity due to the formation of long-lived edge {\it breathers} -- spatially localized nonlinear states with oscillating intensity along the propagation distance. For stronger nonlinearities, localized edge states are not formed for a range of phases, destroying the localization of the intensity correlation. Our results highlight the interplay of band structure, initial state, and nonlinearity influencing transport and intensity correlations.
\end{abstract}
\maketitle

\section{Introduction}\label{intro}
%{\it Introduction.}

Transport and localization in discrete lattices reveal various intriguing phenomena influenced by system parameters, gauge fields, interactions, and nonlinearity. In recent years, photonic %Photonic 
and ultra-cold atomic systems~\cite{christodoulides2003discretizing, bloch2012quantum} {have %recently 
found widespread applications}
%have become natural platforms 
for predicting and observing such effects, 
including quantum walk~\cite{perets2008realization, karski2009quantum}, 
Anderson localization~\cite{schwartz2007transport, billy2008direct},
two-body physics~\cite{corrielli2013fractional, preiss2015strongly, mukherjee2016observation}, %lahini2012quantum
and topologically protected transport~\cite{ozawa2019topological, smirnova2020nonlinear, aidelsburger2015measuring}. 
{Specifically, photonic lattices consisting of evanescently coupled waveguides have become a natural platform for quantum simulations %~\cite{feynman1982simulating}
due to the recent advancements in fabrication and probing methods.} %{\color{red} Should talk about non Hermitian physics also.}

A rich variety of physical phenomena can be accessed by measuring intensity correlations~\cite{paul1986interference, glauber1963photon}, as demonstrated 
by Hanbury Brown and Twiss (HBT) in their groundbreaking experiments on stellar interferometry
~\cite{hanbury1956test} and photon bunching~\cite{brown1956correlation}. %{\color{red}, phenomena that cannot be accessed through mere intensity measurements}. 
{ Detecting correlations in signal intensities, rather than amplitudes, has %since then 
inspired novel measurement techniques across various fields.
For example, probing %temporal 
intensity correlations is a powerful tool for characterizing light sources and observing non-classical behaviors through antibunching~\cite{lounis2005single}. % {\color{red}( the nature of interactions between propagating particles [current 25])}.
In addition, the quantum interference and statistical properties of bosons and fermions can be readily distinguished using HBT intensity interferometry \cite{henny1999fermionic, jeltes2007comparison}.
% {\color{red} (provides information on the space-time geometry of collision, which depends on the wave mechanics of the produced particles)}. 
Intensity correlation experiments also serve as an important tool in high-energy particle physics for understanding the dynamical evolution of heavy-ion collisions \cite{boggild1993identified, baym1998physics}.}
In optics, it is of great interest to understand how two-point correlations manifests when the optical fields propagate through an engineered medium rather than free space~\cite{bromberg2009quantum}. In periodic photonic structures, light transport and intensity correlations can be strongly influenced by the photonic bands and the input state.
Additionally, the quantum nature of light~\cite{peruzzo2010quantum, klauck2021photonic} and nonlinear %light-matter 
interactions~\cite{bromberg2010hanbury} can give rise to highly non-intuitive effects.

In this work, we measure spatial intensity correlations in a photonic Su-Schrieffer-Heeger (SSH) lattice \cite{su1979solitons, malkova2009observation, weimann2017topologically, roberts2022topological} {using phase-tunable coherent states. Such a one-dimensional waveguide array with bipartite weak and strong couplings is an elegant system for exploring topology, nonlinear interactions~\cite{dobrykh2018nonlinear, xia2020nontrivial, li2023interaction}, and quantum photonics~\cite{blanco2018topological}.} 
Intensity correlation measured in the linear regime is used to obtain the dynamics of two indistinguishable bosons. Specifically, we experimentally simulate two-body bound states in the continuum (BICs) %formed by the two correlated bosons.
that are robust against disorders.
We note that BICs \cite{von1929merkwurdige, hsu2016bound, plotnik2011experimental, PhysRevLett.111.220403, cerjan2020observation, friedrich1985interfering, robnik1986simple} are localized states embedded in a band of delocalized states that have been mostly explored in the single-particle regime; see also Ref.~\cite{zhang2012bound, della2014floquet, di2016two, gorlach2017topological, sugimoto2023many, Zhang2023}. 
%{\color{red}Photonic platforms are particularly important for exploiting the intriguing properties of BICs along with their applications in highly efficient light confinement, sensing, and polarization conversion~\cite{yu2019photonic, azzam2021photonic, kang2023applications}.}
%
In our case, due to particle statistics,  two non-interacting bosons can occupy the topological edge states of the SSH lattice and constitute localized bound states in the continuum of a two-body bulk band~\footnote{Unlike bound states, such as {\it doublons,} particle interactions are not necessary to form BICs.}. We also construct two-body semi-localized states associated with one particle on the edge of the lattice and one in the bulk.
{A generalized intensity correlation is then introduced to experimentally simulate the dynamics of two identical spinless fermions for which BICs do not exist because of the Pauli exclusion principle.} 
%

%Next, we measure intensity correlations in the presence of optical Kerr nonlinearity. %and probe long-lived discrete breathers~\cite{flach1998discrete, kopidakis2000discrete, mandelik2003observation, trias2000discrete, johansson2023topological}. 
%{\color{red} (We note that such a nonlinearity is equivalent, in a second-quantized formalism, to having interactions between photons~\cite{Kerr_interactions}).}
%
{Next, we measure intensity correlations in the presence of optical Kerr nonlinearity, which is equivalent to effective interactions among photons mediated by the ambient medium~\cite{haus1989quantum}.}
{ At weak nonlinearities, the localization of intensity correlation persists due to the formation of long-lived discrete breathers and solitons on the edge of the lattice. When the nonlinear strength is further increased, the peak at the corner of the correlation matrix disappears.}
Breathers~\cite{flach1998discrete, kopidakis2000discrete, mandelik2003observation, trias2000discrete, johansson2023topological} are spatially localized nonlinear states that are periodic along the propagation distance (i.e., the analogous `time,' in our case). 
%{\color{red} They appear in various systems including Bose-Einstein condensates, superconducting materials, Josephson junctions, and photonic structures.
The periodic oscillations of intensity make breathers distinct from solitons, which are shape-preserving nonlinear states~\cite{barthelemy1985propagation, christodoulides1988discrete, segev1992spatial, eisenberg1998discrete, suchkov2016nonlinear, Pernet2022}. 
In addition, unlike Floquet solitons~\cite{mukherjee2020observation} appearing in periodically driven systems, breathers are formed in static systems still exhibiting periodic dynamics.

{The paper is organized as follows. In Sec.~\ref{intensity_corre}, we discuss our experimental platform and define intensity correlations along with the procedure for its measurement. In Sec.~\ref{mappingbosonic}, we discuss how the evolution of two indistinguishable photons (bosons) through a waveguide lattice can be experimentally simulated by the measured intensity correlation.
In Sec.~\ref{Two_body_BIC}, we present the two-body spectrum of a finite SSH chain supporting two-body edge BICs, and we show our experimental results. 
In Sec.~\ref{Two_fermion}, we discuss how the dynamics of two identical fermions can be constructed using a generalized
intensity correlation. Then, in Sec.~\ref{NLeffects}, we consider the nonlinear transport of light, and we observe long-lived edge breathers.
Finally, we highlight the significance of this work and its future implications in Sec.~\ref{conl}.
}

%\section{Results} \label{results}
\section{Intensity correlation in \\photonic lattices} 
\label{intensity_corre}

%{\it Intensity correlation in photonic lattices.}
The scalar-paraxial transport of optical fields in a photonic lattice can be described by the discrete nonlinear Schr{\"o}dinger equation
%$i\frac{\partial}{\partial z} \psi_j(z)=\sum H_{jj'}^{\text{lin}} \psi_{j'} -|\psi_{j}|^2\psi_{j}, $
\begin{eqnarray} \label{SE}
i\frac{\partial}{\partial z} \psi_j(z)=\sum H_{jj'}^{\text{lin}} \psi_{j'} -|\psi_{j}|^2\psi_{j} \; , \label{eq1}
\end{eqnarray} 
where the propagation distance plays the role of time $(z\leftrightarrow t)$, $\psi_j$ is proportional to the slowly-varying complex amplitude of the optical field at the $j$th waveguide, and $\hat{H}^{\text{lin}}$ is the linear tight-binding Hamiltonian containing the evanescent couplings and onsite energies (i.e., propagation constants). 
In the absence of optical loss, the total energy and the renormalized power ($g\equiv\sum_j |\psi_{j}|^2$) are conserved.
Here, $g$ quantifies the strength of the {optical Kerr} nonlinearity, which is negligible at low optical powers. 

\begin{figure}[t!]
    %\centering
    \includegraphics[width=1\linewidth]{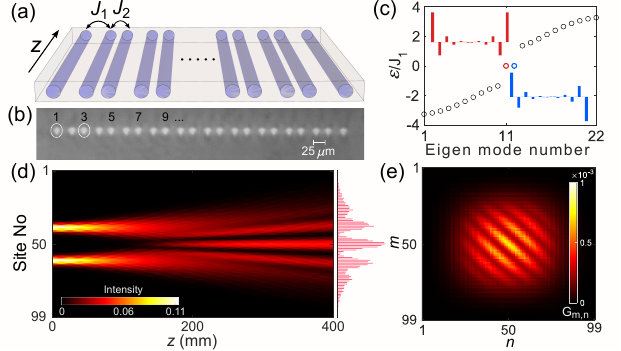}
    \caption{(a) Sketch of a photonic SSH lattice terminated with weak couplings $J_1\!<\!J_2$. The couplings are experimentally controlled by varying the inter-site spacing. (b)
Micrograph of the cross-section of a fs laser-written SSH lattice. (c) Spectrum of the SSH lattice (open circles). Notice the appearance of the edge modes in the band gap. (Insets) amplitude distribution of the edge modes. 
(d) { Numerical results show that} Young's double slit-type interference fringes are created during light propagation { in the linear regime}. These fringes disappear if the relative phase of the initial states fluctuates rapidly. (e) In a phase-averaged detection, the intensity correlation $G_{m, n}(z\!=\!400 \,{\text{mm}})$ exhibits fringes. 
    }
    \label{fig_1}
\end{figure}

Figure~\ref{fig_1}(b) shows a schematic of 
a photonic SSH lattice with dimerized nearest-neighbor couplings $J_{1, 2}$ and equal onsite energies.
For a finite SSH lattice, 
topological modes appear on the boundaries when the lattice is terminated with weak couplings. 
In experiments, an SSH lattice consisting of $22$ sites is created using fs laser-writing~\cite{davis1996writing, szameit2010discrete} in an $80$-mm-long borosilicate (BK7) glass substrate; see Fig.~\ref{fig_1}(b). 
At $1030$ nm wavelength of light, the couplings were measured to be $J_1\!=\!0.042$~mm$^{-1}$ and $J_2\!=\!0.096$~mm$^{-1}$ -- these couplings have been used in all calculations unless otherwise mentioned. 
The band structure of the lattice is presented in Fig.~\ref{fig_1}(c), where the insets show the edge mode amplitude profiles.
%and the amplitude distribution of the zero-energy topological modes are shown in Fig.~\ref{fig_1}(h).

%\subsection{Intensity correlation}
Consider the propagation of two broad and equally intense Gaussian 
wave packets with a tunable relative global phase $\phi$ in a photonic SSH lattice; Fig.~\ref{fig_1}(a, b).
After some propagation, these two wave packets interfere and produce discrete interference fringes, %; see Fig.~\ref{fig_1}(d) for $\phi=0$.
as shown in Fig.~\ref{fig_1}(d) for $\phi\!=\!0$. {A large system size of $99$ waveguides is considered such that light does not reach the edge of the lattice during the maximal propagation distance considered here.}
Averaging the intensity distribution over initial relative phases completely washes out these interference fringes. %{\color{SMblue}Fig.~\ref{fig_1}(d).}
However, the fact that interference has taken place can still be inferred {%measuring 
from the following spatial intensity} correlations at a propagation distance $z$
%$G_{m, n}(z)\!=\! \frac{1}{2\pi}\int_{0}^{2\pi}\text{d}\phi\, I_m(\phi, z)I_n(\phi, z),$
\begin{eqnarray} 
  G_{m, n}(z)\!=\! \frac{1}{2\pi}\int_{0}^{2\pi}\text{d}\phi\, I_m(\phi, z)I_n(\phi, z) \,, \label{Gmn}
\end{eqnarray}
where $I_m\!=\!|\psi_m (z)|^2/\sum_n |\psi_n (z)|^2$ is the normalized intensity at the $m$-th site; %as shown in 
see  Fig.~\ref{fig_1}(e).  
Note that such broad initial states excite the low-energy eigenstates, and hence, $G_{m, n}$ does not carry the full information of the band structure in this case.
The width of the wave packets can be reduced to make the influence of the band structure apparent.
In fact, from now on we will consider equally intense single-site excitation near the edge waveguides with a tunable relative phase.

In experiments, we launch temporally synchronized laser pulses at $1030$ nm wavelength into two desired lattice sites {
of the photonic lattice.
%with a tunable relative phase,
Before coupling to the lattice, one of the pulses travels through a $1$ mm thick glass plate that can be rotated to vary the relative phase {$\phi$ from $0$ to $2\pi$ with a step size of $\pi/100$.} 
%standard deviation of $\Delta\phi\!=\!\pi/30$.
The pulse energy and phase of the input excitation are precisely controlled,} 
and the output intensity distributions are recorded using a CMOS camera; {
%~\cite{Suppmat}.
%Method Section.
see Appendix~\ref{fab_char}.} Note that, unlike intensity profiles, the intensity correlation is less sensitive to low-frequency phase fluctuations. 
{ In the following sections, we discuss how the correlation matrix can be used to construct the dynamics of two non-interacting particles. Additionally, the influence of optical Kerr nonlinearity and the appearance of discrete breathers are discussed in Section~\ref{NLeffects}.}

{ 
\section{Construction of two-body dynamics}
\label{mappingbosonic}}

%{\color{SMblue}{\it Construction of two-body dynamics.}} 
The evolution of certain two-body quantum states {of light through a photonic lattice} 
can be captured by intensity correlations~\cite{bromberg2009quantum, keil2010photon}. Indeed, the quantum walk of two indistinguishable bosons (photons) can be emulated by probing the intensity correlation {in Eq.~\eqref{Gmn}} 
in the linear regime.
{ 
To establish the mathematical connection, we note that the $z$-evolution of the bosonic creation operator at site $q$ is obtained by integrating the Heisenberg equation: %~\cite{Suppmat}:
$\hat{a}^{\dagger}_{q}(z) = \sum_{m}U_{qm}(z)\hat{a}^{\dagger}_{m}$, where $\hat{U}(z) = \exp{(iz\hat{H}^{\text{lin}})}$ is the unitary evolution matrix, and $\hat{H}^{\text{lin}}$ is the lattice Hamiltonian; see also Appendix~\ref{secA}.  
The modulus squared of the two-photon wave function at a propagation distance $z$ for an input state $\ket{p,q} \!=\! \hat{a}^{\dagger}_{p}\hat{a}^{\dagger}_{q}\ket{0}$ initially coupled at waveguide $p$ and $q\ne p$ is
%The modulus square of the two-photon wave function is 
then given by 
\begin{eqnarray} \label{psimn2}
|\Psi_{m,n}(z)|^2\!=\!\frac{1}{2}|U_{mp}(z)U_{nq}(z)+U_{mq}(z)U_{np}(z)|^2 \,.
\end{eqnarray}
For the sake of presentation, we show $|\Psi_{m,n}(z)|^2$ in a square grid, which means that both indices, $m$ and $n$, run from $1$ to $N$, where $N$ is the total number of sites in the lattice.}

{
To construct $|\Psi_{m,n}(z)|^2$ shown in Eq.~\eqref{psimn2} using the transport of coherent states of light,
we consider an initial state $\Psi_p(z\!=\!0)\!=\!1/\sqrt{2}$ and $\Psi_q(z\!=\!0)\!=\!\exp(i\phi)/\sqrt{2}$ coupled to the lattice sites $p$ and $q$, respectively, with a relative phase $\phi$.
In this case,  the output intensity at the $m$-th site after a propagation through distance $z$ is given by
$I_m(\phi,z) = \frac{1}{2}|U_{mp}(z)+U_{mq}(z)e^{i\phi}|^2$ -- note the appearance of the evolution matrix elements.
The correlation matrix in Eq.~\eqref{Gmn} is then calculated for such a two-site initial state, giving the following mathematical relation between $G_{m,n}$ and $|\Psi_{m,n}(z)|^2$,%~\cite{Suppmat} 
}
\begin{eqnarray}
    |\Psi_{m,n}(z)|^2\!=\! \frac{1}{2}\left[ 4G_{m, n} - I_m^p I_n^p - I_m^q I_n^q \right] \, , \label{TBState}
\end{eqnarray} %% 
where $I_m^{p}$ is the intensity at the $m$-th site for the initial excitation at the $p$-th site only; see Appendix~\ref{secA}.
{ Since $I_m^{p}$ is measured by evolving a single-site initial state, it is independent of the relative phase.
Further, $G_{m,n}$ is obtained by averaging over the relative phase between the two-site initial state, and hence, it does not explicitly depend on the relative phase.}
Evidently, by measuring all three terms on the right-hand side of Eq.~\eqref{TBState}, we obtain $|\Psi_{m,n}|^2,$ the probability of detecting one photon at site $m$ and one photon at site $n$ at the output of the lattice. 
This approach can also be extended for two-dimensional lattices to construct $|\Psi_{m,n}|^2$ from the single particle dynamics in the non-interacting limit.\\
\\

\begin{figure}[t!]
\centering
    \includegraphics[width=1\linewidth]{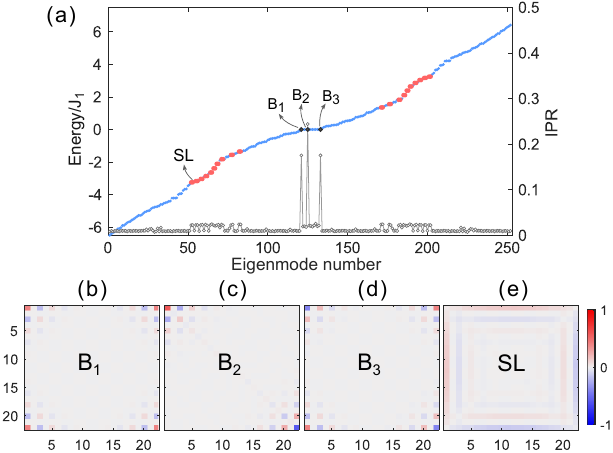}
    \caption{(a) Two-body energy spectrum (filled circles) and the IPR of the eigenstates (open circles) of the SSH lattice consisting of 22 sites and terminated with weak couplings.  Black circles indicate two-body edge BICs (B$_{1-3}$). Here ${J_{1,2}\!=\!\{0.042, 0.096\}}$~mm$^{-1}$. (b-d) and (e) The wave function of the two-body BICs and a semi-localized (SL) state, respectively. %, %are indicated by black circles and their wave functions are shown in (b). Here ${J_{1,2}\!=\!\{0.042, 0.096\}}$~mm$^{-1}$. (e) A semi-localized (SL) state is also shown in (b). 
    }
    \label{fig_2}
\end{figure}

\section{Two-body edge BICs}
\label{Two_body_BIC}
% {\it Two-body edge BICs.}

The two-body energy spectrum of a finite SSH chain terminated with weak couplings is shown in Fig.~\ref{fig_2}(a).
Three bands of delocalized bulk states are indicated in blue, and another two bands of semi-localized states associated with one particle on the edge and one in the bulk are shown in red.
In the same Fig.~\ref{fig_2}(a), we quantify the localization of each eigenstate $\Phi_{m,n}$ by plotting the inverse participation ratio, ${\text{IPR}}\!\equiv\! \sum |\Phi_{m,n}|^4/ \left(\sum|\Phi_{m,n}|^2 \right)^2$.
The three most localized states ($B_1, B_2$ and $B_3$, indicated by the black circles) located at the edge of the lattice are almost degenerate at zero energy.
These states correspond to a finite-size mixing of edge modes where two particles live on the same edge or on opposite edges \footnote{Denoting the SSH single-particle edge modes by $|L\rangle$ and $|R\rangle$, the two particle states $|L\rangle \otimes |L\rangle$, $|R\rangle \otimes |R\rangle$ and \mbox{$(|L\rangle \otimes |R\rangle + |R\rangle \otimes |L\rangle)/\sqrt{2}$} are eigenstates of the two-particle non-interacting problem in the thermodynamic limit}.
Importantly, these localized two-body edge modes are in resonance with the delocalized bulk modes, and hence, they act as  bound states in the continuum. 
{The amplitude distributions of the three BICs and one of the semi-localized states are presented in Figs.~\ref{fig_2}(b-d) and (e), respectively.}
From the amplitude distributions in %the insets of 
Fig.~\ref{fig_2}(b-d), we can infer that the edge BICs correspond to
a significant probability of finding one particle at site $1$ and one at site $3$. 
In other words, the BICs can be experimentally probed by preparing 
the two particles initially located at sites $1$ and $3$,  i.e., $\Psi_{1,3}^{\text{in}}=\ket{1,3}$, as described below.
Note that these BICs are exponentially localized to the edges of the lattice. %,  
More importantly, they exist even in the presence of random diagonal and off-diagonal disorder ($\lesssim\!J_1/2$),   
showing remarkable robustness. %;  %Supplementary Material. 
{Indeed, the eigenvalues of the BICs are not affected by a small coupling disorder. The diagonal disorder can make the BICs non-degenerate; however, they still remain
localized and embedded in the two-body bulk band as
long as the disorder strength is $\lesssim J_1/2$; see Appendix~\ref{bic_robust}.
}

\begin{figure}[hbt]
\centering
\includegraphics[width=1\linewidth]{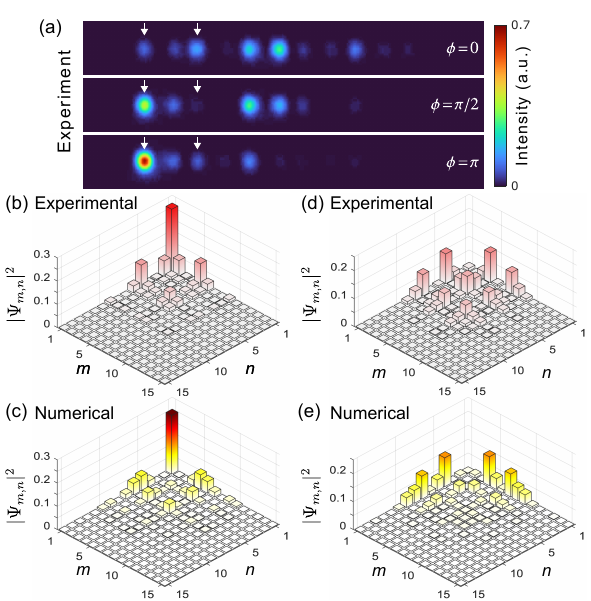}
    \caption{(a) Measured intensity patterns at the output (${z\!=\!80}$ mm) of the SSH array for three different relative phases ${\phi\!=\!0, \, \pi/2}$ and $\pi$. White arrows indicate sites $1$ and $3$, where the light was coupled initially. (b,c)  Experimentally obtained ${|\Psi_{m,n} (z=80 \, \text{mm})|^2}$ agrees with numerical results. 
(d,e) Excitation of the red-colored semi-localized states in Fig.~\ref{fig_2}(a).
}
    \label{fig_2b}
\end{figure}

We launch coherent states of light at waveguides $1$ and $3$ of the SSH lattice and measure the output intensities at $z\!=\!80$ mm as a function of the relative phase $\phi$.
By tuning the relative phase from $0$ to $2\pi$, 
we can control the overlap of the initial state with the single-particle states in this case. Due to the amplitude and phase profile of the topological edge modes (see Fig.~\ref{fig_1}(h)), their overlap with the input state is maximal (minimal) for $\phi\!=\! \pi (0)$.
Experimentally measured output intensity patterns for $\phi\!=\!0, \; \pi/2$, and $\pi$ are shown in Fig.~\ref{fig_2b}(a). 
Note that the optical power located at the edge is maximal (minimal) for the initial state with $\phi\!=\!\pi$ ($0$) relative phase, as would be expected; see also 
% Supplementary 
% Figs.~S5(a-c).
\ref{fig_lin_dynamics}(a-c).

By performing the phase-averaged measurement, we obtain $|\Psi_{m,n}|^2$, %as shown in Fig.~\ref{fig_2}(c), 
which is in excellent agreement with our numerical results; see Figs.~\ref{fig_2b}(b,c).
%
%It is evident 
Notice that the probability of detecting both particles at the edge site $1$, i.e., $|\Psi_{1,1}|^2$, is maximal. 
For $J_2\!>\!J_1$, $|\Psi_{1,1}|^2$ first oscillates as a function of $z$, and then slowly approaches a nonzero value that is determined by $J_2/J_1$; see 
% Supplementary 
Fig.~\ref{BIC_sup}.
In addition to the BICs, the above input state overlaps with the red-colored semi-localized states shown in the spectrum, Fig.~\ref{fig_2}(a). As a result, we notice non-zero elements $|\Psi_{1,n}|^2$ and $|\Psi_{m,1}|^2$ ($m, n \ne 1$) in Figs.~\ref{fig_2b}(b,c). However, the prominent peak at \{1, 1\} is due to the BICs only. 
{Evidently, our experimental simulation predicts the evolution of photon pairs in waveguide arrays -- if two indistinguishable photons are launched at sites 1 and 3 of the SSH array, they will be bunched and will emit together from the edge with maximal probability.}

To gauge the significance of the initial state and to prove that the semi-localized states do not cause a prominent peak at \{1, 1\}, we performed another correlation measurement by coupling light to sites $1$ and $2$ of the SSH lattice -- the analogous two-body input state is $\Psi_{1,2}^{\text{in}}=\ket{1,2}$. In this case, the semi-localized states in the spectrum Fig.~\ref{fig_2}(a) are efficiently excited, 
but the excitation of the BICs is insignificant; see 
% Supplementary
Fig.~\ref{BIC_sup}(b).
Hence, one photon penetrates the bulk and the other one lives on the boundary; however, $\Psi_{1,1}$ is insignificant, as shown in Fig.~\ref{fig_2b}(d,e).
We note that these BICs do not exist for $J_1\!\ge \!J_2$ (trivial case), which was experimentally verified for a one-dimensional lattice with $J_1\!=\!J_2$.

\begin{figure}[hbt]
\centering
    \includegraphics[width=1\linewidth]{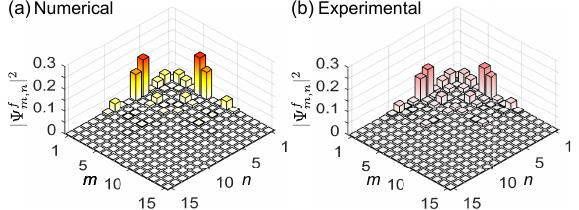}
    \caption{(a,b) Numerically and experimentally obtained ${|\Psi^f_{m,n} (z=80 \, \text{mm})|^2}$ for two identical fermions, respectively. }
    \label{fig_fermion}
\end{figure}

{ 
\section{Two-fermion state}
\label{Two_fermion}

%{\it Two-fermion state.}
In this section, we show that the dynamics of two identical fermions can be constructed using a generalized  intensity correlation.
%
%\ref{mappingbosonic}
In the case of two identical spinless fermions, it is straightforward to show that the modulus squared of the two-body wave function in Eq.~\eqref{psimn2} is modified as
\begin{align}
|\Psi^f_{mn}(z)|^2 = \frac{1}{2}|U_{mp}(z)U_{nq}(z)-U_{mq}(z)U_{np}(z)|^2 . \label{psimn2Fermion}
\end{align}
Here, the superscript $f$ indicates that we are considering two-body fermionic states, and the negative sign in Eq.~\eqref{psimn2Fermion} arises due to fermionic statistics, which leads to destructive interference under particle exchange. 
To experimentally simulate the above two-fermion correlation, we
%We now 
define a generalized intensity correlation 
$$G_{mn}(z, \theta)\!=\!\frac{1}{2\pi}\int_{0}^{2\pi}\text{d}\phi\, I_m(\phi, z)I_n(\phi+\theta, z)\, ,$$ 
where $\theta\!=\!0$ corresponds to Eq.~\eqref{Gmn}. Interestingly, the dynamics of two identical spinless fermions can be simulated by measuring the above generalized intensity correlation for $\theta\!=\!\pi$; see Appendix~\ref{secA}.
%supplementary material~\cite{Suppmat}. 
Considering the two fermions initially located at sites 1 and 3, we obtained the modulus squared of the wave function $|\Psi^f_{m,n}|^2$ at $z\!=\!80$~mm; see Fig.~\ref{fig_fermion}.  
In Fig.~\ref{fig_fermion}(b), the experimentally constructed probability of finding both fermionic particles together (i.e., the elements along the main diagonal, $m\!=\!n$) is negligible within experimental signal-to-noise ratio errors, as expected from the Pauli exclusion principle. 
Unlike the bosonic case in Figs.~\ref{fig_2b}(b,c), no peak is observed at $\{1,1\}$, demonstrating that no BIC can form in this case.}

\section{Nonlinear effects}\label{NLeffects}

\begin{figure}[hbt]
\centering
    \includegraphics[width=1\linewidth]{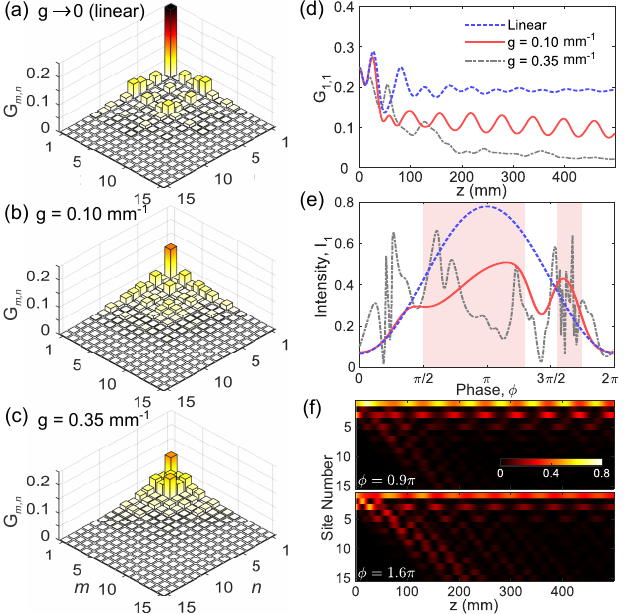}
    \caption{(a-c) Numerically obtained intensity correlations ${G_{m,n} (z\!=\!80 \,{\text{mm}})}$ in the linear, weak, and strong nonlinear regimes for input states coupled to site 1 and 3. (d) The variation of ${G_{1,1}}$ with propagation distance. (e) Intensity at waveguide 1 at ${z\!=\!80 \,{\text{mm}}}$ as a function of $\phi$ for three different nonlinearities.
The reddish shade is a guide to the eye, indicating the phase values for which long-lived edge breathers are formed at ${g\!=\!0.1}$ mm$^{-1}$ (red curve).
(f) Dynamics of the two-site input state for ${g\!=\!0.1}$~mm$^{-1}$, confirming the formation of long-lived edge breathers at ${\phi\!=\! 0.9\pi}$ and ${1.6\pi}$. Only $15$ lattice sites are shown here.
}  
    \label{fig_3}
\end{figure}

We now consider nonlinear 
{transport of light waves coupled to 
sites $1$ and $3$ of the photonic SSH} lattices.
In the case of homogeneous couplings $J_1\!=\!J_2$, intensity correlations exhibit a delocalization-to- localization feature with increasing nonlinearity (see 
% Supplementary 
% Fig.~S8
Fig.~\ref{int_corre}), because the self-focusing nonlinearity reduces the spreading of light irrespective of the relative phases.
On the contrary, $G_{m,n}$ for the SSH chain shows a localization-to-delocalization behavior, Fig.~\ref{fig_3}(a-c) -- the peak of $G_{m,n}$ at $\{1, 1\}$ reduces with increasing nonlinearity.
To highlight the dynamics, we plot the variation of $G_{1, 1}$ with $z$ in Fig.~\ref{fig_3}(d). 
In the linear case, $G_{1, 1}$ reaches a fixed value  after some initial transient dynamics.
For $g\!=\!0.1$ mm$^{-1}$, $G_{11}(z)$ oscillates around a mean value that decays slowly, whereas at a larger nonlinear strength $g\!=\!0.35$ mm$^{-1}$, $G_{1, 1}(z)$ decays fast along $z$.
In other words, the peak of $G_{m,n}$ at $\{1, 1\}$ persists for weak nonlinear strengths; however,
it is destroyed at a larger nonlinearity.
Note that $G_{1, 1}$ is the phase average of intensity squared at waveguide 1, i.e., $G_{1, 1}\!=\!\langle I_1^2 \rangle$. 
Hence, to understand the influence of nonlinearity in Figs.~\ref{fig_3}(a-c), we plot $I_1$ as a function of $\phi$ for the three different nonlinear strengths; see Fig.~\ref{fig_3}(e). 
In the linear regime, $I_1$ exhibits one peak at $\phi\!=\!\pi$ because the initial state overlaps maximally with the topological edge modes in this case.
As $\phi$ deviates from $\pi$, the bulk modes are excited more, and hence, $I_1$ reduces; see also 
% Supplementary 
% Figs.~S5(a-c). 
Figs.~\ref{fig_lin_dynamics}(a-c). 
Evidently, the phase average of $I_1$ (and hence, $I_1^2$) is significant in the linear case, giving rise to the prominent peak in Fig.~\ref{fig_3}(a) at the corner, $\{1, 1\}$.

\begin{figure}[hbt]
\centering
    \includegraphics[width=1\linewidth]{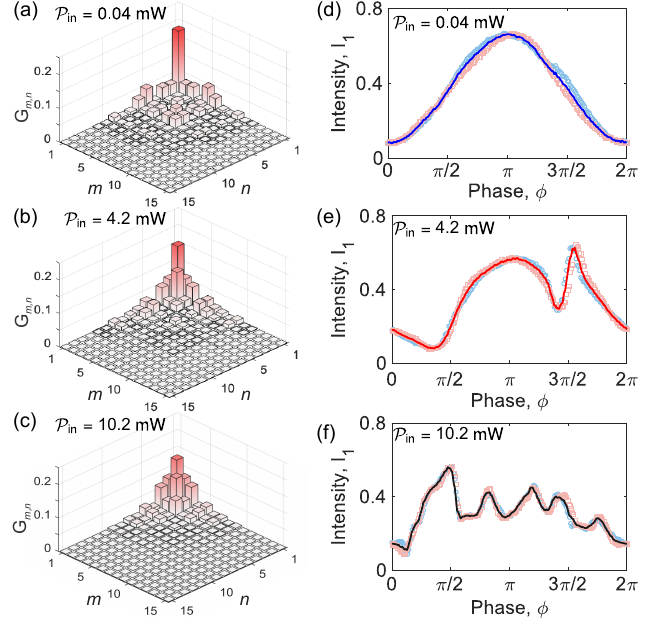}
    \caption{(a-c) Experimentally measured intensity correlation ${G_{m,n} (z\!=\!80 \,{\text{mm}})}$ at three different input powers ${\mathcal{P}_{\text{in}}}$ indicated on each image. The field of view (${m \times n}$) is smaller than the actual experiments. (d-f) Measured intensity at waveguide 1 as a function of $\phi$ associated with (a-c). Due to the formation of long-lived edge breathers, two prominent peaks are observed in (e). Data sets for two independent measurements and their mean are shown by red squares, %and 
    blue circles, and a solid line, respectively. 
    }
    \label{fig_4}
\end{figure}
The oscillatory behavior of $G_{1,1}(z)$ in the weak nonlinear regime %(Fig.~\ref{fig_3}(d)) 
indicates the possible formation of localized nonlinear edge states {\it over a range of relative phases}.
Indeed, the nonlinear dynamics in 
% Supplementary 
% S6(b-d)
Figs.~\ref{fig_NL_dynamics}(b-d) 
and Supplementary Movie 2 confirm the formation of long-lived edge breathers. %and solitons at the edge of the lattice.
The two peaks observed in the variation of $I_1(\phi)$ (see the red curve in Fig.~\ref{fig_3}(e)), near $\phi\!=\! (0.9 \pm 0.4)\pi$ and $(1.65 \pm 0.1)\pi$, are the signatures of the edge breathers as demonstrated experimentally below.
The nonlinear dynamics of the two-site input state is presented in Fig.~\ref{fig_3}(f), showing the edge breathers at $\phi=0.9\pi$ and $\phi=1.6\pi$.
As mentioned before, a discrete breather is a nonlinear state localized along the lattice axis and periodic along $z$. 
{In our case, the photonic SSH lattice is described by a static (i.e., $z$-independent) and linear Hamiltonian $\hat{H}^{\text{lin}}$. 
When a breather is formed, the nonlinearity locally modifies the refractive index of the waveguides in a $z$-periodic manner. In other words, the total Hamiltonian $\hat{H}_t\!=\!\hat{H}^{\text{lin}}+\hat{H}^{\text{nl}}$ becomes periodic in $z$ with a periodicity of the breather, i.e., $\hat{H}_t(z+z_0)=\hat{H}_t(z)$, where $z_0$ is the period, and $\hat{H}^{\text{nl}}$ denotes the nonlinear part of the Hamiltonian associated with Eq.~\eqref{eq1}.}
We can then use the Floquet self-consistency algorithm~\cite{lumer2013self} to look for a true breather solution and its quasienergy; see 
% Supplementary 
% Fig.~S9.
Fig.~\ref{breather_fig}.
Now, for a localized nonlinear state with $\approx \!58$ mm period, as in the case of Fig.~\ref{fig_3}(f), the Floquet quasienergy spectrum would be gapless due to band folding; see also Appendix~\ref{DB}. %~\cite{Suppmat}.
In other words, {%these 
the long-lived edge breathers in Fig.~\ref{fig_3}(f)}
are in resonance with the linear modes, causing their finite lifetime. 
We also notice the formation of edge solitons at specific phases, e.g., at $\phi=0.3\pi$ in 
% Supplementary 
% Fig.~S6(a). 
Fig.~\ref{fig_NL_dynamics}(a). 
Since these edge solitons require relatively low power, most of the light was observed to penetrate in the bulk.

In the case of $g\!=\!0.35$ mm$^{-1}$, the dynamics is more sensitive to relative phase $\phi$; see 
% Supplementary 
% Figs.~S6(e-h)
Figs.~\ref{fig_NL_dynamics}(e-h) 
and Supplementary Movie 3.
Notice the rapid changes in $I_1(\phi)$, particularly near $\phi \!=\! \pm \pi/2$. %As shown in Supplementary Movie 3, 
Long-lived localized nonlinear edge modes are not formed for a wide range of $\phi$, in this case. As a result, the peak of $G_{m,n}$ at $\{1, 1\}$ is almost destroyed,  Fig.~\ref{fig_3}(c). 
On a separate note, in the limiting case of $g \gg J_2$, the initial state will be trapped, hence, the localization of $G_{m,n}$ at $\{1, 1\}$ can be restored.

To probe the above nonlinear effects, we employ temporally stretched and down-chirped laser pulses.
The unwanted effect of self-phase modulation, % 
chromatic dispersion and multi-photon absorption were found to be negligible in our experiments, and the
nonlinear strength $g$ was estimated to be %calibrated by characterizing two-waveguide devices and was found to be 
$g=0.045  \, \mathcal{P}_{\text{in}}$  mm$^{-1}$mW$^{-1}$. %~\cite{Suppmat}.
The intensity correlations $G_{m,n}(z\!=\!80\, {\text{mm}})$ for three different $\mathcal{P}_{\text{in}}$ are shown in Figs.~\ref{fig_4}(a-c), which are in good agreement with Figs.~\ref{fig_3}(a-c).
The associated variations of $I_1$ with $\phi$ are shown in Figs.~\ref{fig_4}(d-f), respectively. % for two independent measurements. 
Notice a single peak in the linear case and a double peak in the weak nonlinear regime; Fig.~\ref{fig_4}(d,e) -- here two independent data sets are shown. In our experiments, %the second peak 
two peaks in $I_1(\phi)$ were observed for a range of power ($2.1$~mW to $4.2$~mW), clearly proving the existence of the long-lived edge breathers. When nonlinearity was further increased, more peaks started appearing -- Fig.~\ref{fig_4}(f) shows the case of $\mathcal{P}_{\text{in}}\!=\!10.2$~mW. 
It should be noted the front and rare tails of the laser pulse behave linearly, giving rise to small deviations between experiments and numerics. However, by inspecting Fig.~\ref{fig_3} and Fig.~\ref{fig_4}, we notice a good qualitative agreement.

\section{Conclusions}\label{conl}
%{\it Conclusions.}

We have experimentally and numerically studied light transport and intensity correlations in a finite SSH lattice. % 
Intensity correlations, measured in the linear optical regime, are used to experimentally emulate the dynamics of two identical bosons as well as fermions in the lattice. 
{
This work opens an exciting route towards the investigation of BICs %and higher-order BICs~\cite{} 
in the context of two-particle physics. %
Specifically, two-body BICs can find potential applications in the robust generation and transport of correlated photons.
Our methodology can also be employed to elucidate the existence and behavior of two-particle BICs in two-dimensional lattices, %and models, 
e.g., those describing higher-order topological insulators \cite{cerjan2020observation}, where distinct protection mechanisms and topological properties will manifest.
Additionally, this work can be extended to `time'-modulated %two-dimensional 
lattices~\cite{tschernig2021topological} to understand how topological protection is carried over for multi-particle quantum states, such as indistinguishable photons and NOON states.}
In the nonlinear regime, intensity correlations enable us to demonstrate long-lived discrete edge breathers.
Unlike the traditional single-site excitation experiments~\cite{ mukherjee2020observation, mukherjee2021observation, guo2020weakly}, coupling light into two lattice sites with a tunable phase allows us to selectively excite the edge and bulk states and then probe nonlinear dynamics. It also allows us to access relatively higher nonlinearities by injecting more light in the lattice, which is usually limited by the laser-induced material damage threshold. 
{ 
Precise multi-site input state preparation, together with the flexibility in realizing complex lattice Hamiltonians, make our platform suitable for observing various intriguing phenomena, such as the formation and collision of nonlinear discrete solitons, breathers, and multi-breathers~\cite{flach1998discrete, bountis2000multibreathers, akhmediev2016roadmap}.} 
Furthermore, it opens the intriguing possibility of investigating multi-particle {(two or more)} correlations when light is launched in multiple sites with appropriate phase patterns.

\begin{acknowledgments}
%{\it Acknowledgments.$-$} \par %delete if not applicable))
S.M.~gratefully acknowledges %funding
support from the Indian Institute of Science (IISc) through a start-up grant, and Infosys Foundation, Bangalore. S.M.~also thanks the Indian Space Research Organization (ISRO) and Science and Engineering Research Board (SERB) for funding through the ISRO-IISc Space Technology Cell (code no ISTC/PPH/SM/465) and the Start-up Research Grant (file no SRG/2022/002062), respectively. T.S.~and R.H.~thank IISc for scholarships through the Integrated PhD program. 
M.D.L.~acknowledges support by the Italian Ministry of University and Research via the Rita Levi-Montalcini program. D.S.~thanks SERB, India, for funding through Project No. JBR/2020/000043.
\end{acknowledgments}

% References
\medskip

\bigskip

\appendix

        \setcounter{equation}{0}
        \renewcommand{\theequation}{A\arabic{equation}}%
        \setcounter{figure}{0}
        \renewcommand{\thefigure}{A\arabic{figure}}%
         \setcounter{section}{0}
         \renewcommand{\thesection}{\Alph{section}}%
        % }

\section*{{Appendix} \
\vspace*{-0.0cm}}
%\twocolumngrid

%%%%%%%%%%%%%%%%%%%%%%%%%%%
% Derivation of Eq 3
\section{Derivation of Eq.~\eqref{TBState} \label{secA}}

% {\section{More details on mapping of correlations \label{secA}}}
{\it {\color{TriClr}Quantum correlations $-$}}
In general, a tight binding Hamiltonian in the second quantization notation can be written as
\begin{eqnarray}
\hat{\mathcal{H}} = \dfrac{\hbar c}{n_0}\sum_{m,n=1}^{N}H^{\text{lin}}_{nm}\creop{m}\aniop{n} \, , \label{seq1}
\end{eqnarray}
where $\creop{}$ ($\aniop{}$) is the bosonic creation (annihilation) operator, $\hbar$ is the reduced Planck constant, $c$ is the speed of light in vacuum, $n_0$ is the refractive index of the medium, and $H^{\text{lin}}_{nm}$ is the coupling strength between $n$-th and $m$-th sites.
In Heisenberg picture, the evolution equation of the creation operator is given by
\begin{align}
\dd{\creop{q}(t)}{t} &= -\dfrac{i}{\hbar}\bs{\creop{q}(t),\hat{\mathcal{H}}(t)} , \nonumber \\
\dd{\creop{q}(t)}{t} &= -\dfrac{i}{\hbar}\hat{U}^{\dagger}(t)\bs{\creop{q},\hat{\mathcal{H}}}\hat{U}(t), \nonumber\\
\dd{\creop{q}(z)}{z} &= i\sum_{m}H^{\text{lin}}_{qm}\creop{m}(z)  ,\label{seq2}
\end{align}
where $\hat{\mathcal{H}}(t)\!=\!\hat{U}^{\dagger}(t)\hat{\mathcal{H}}\hat{U}(t)$ and $z \!=\! {ct}/{n_0}$. Integrating the above evolution equation, we get
\begin{eqnarray}
\creop{q}(z) = \sum_{m}U_{qm}(z)\creop{m}(z\!=\!0), \label{seq3}
\end{eqnarray}
where $\hat{U}(z) = e^{iz\hat{H}^{\text{lin}}}$,  the explicit form of the time evolution of the creation operators.
Now, the photon number correlation for a given two-photon input state initially coupled at waveguide $p$ and $q\ne p$, i.e., $\ket{p,q} = \creop{p}\creop{q}\ket{0}$, at a propagation distance $z$ is 
\begin{align}
\Gamma_{mn}(z) &= \bra{p,q}\creop{m}(z)\creop{n}(z)\aniop{n}(z)\aniop{m}(z)\ket{p,q}  \nonumber \\
&= \sum_{s_1s_2s_3s_4}U_{ms_1}(z)U_{ns_2}(z)U^*_{ns_3}(z)U^*_{m s_4}(z)  \nonumber \\
& \hspace{18mm} \times \bra{p,q}\creop{s_1}\creop{s_2}\aniop{s_3}\aniop{s_4}\ket{p,q} , \nonumber \\
&= |U_{mp}(z)U_{nq}(z)+U_{mq}(z)U_{np}(z)|^2 . \label{seq4}
\end{align}

{\it {\color{TriClr}Intensity correlations $-$}} Now let us consider an initial state $\Psi_p(z\!=\!0)\!=\!1/\sqrt{2}$ and $\Psi_q(z\!=\!0)\!=\!\exp(i\phi)/\sqrt{2}$ coupled to the lattice sites $p$ and $q$, respectively, with a relative phase $\phi$.
%In this case, we denote the output intensity at the $m$-th site after a propagation of $z$ by $I_m(\phi, z)$.
In this case,  the output intensity at the $m$-th site  after a propagation of $z$ is given by
$I_m(\phi,z) = \frac{1}{2}|U_{mp}(z)+U_{mq}(z)e^{i\phi}|^2$.
We define a generalized intensity correlation as
\begin{align}
G_{mn}(z, \theta) &= \frac{1}{2\pi}\int_{0}^{2\pi}\text{d}\phi\, I_m(\phi, z)I_n(\phi+\theta, z) \nonumber  \\
&= \frac{1}{4}\left[|U_{mp}(z)U_{nq}(z)+ e^{i\theta} U_{mq}(z)U_{np}(z)|^2 \right.\nonumber \\
&\hspace{10mm} \left.+ I^p_m(z)I^p_n(z) + I^q_m(z)I^q_n(z) \right] \label{seq5-0}
\end{align}
For capturing bosonic statistics, we consider $\theta=0$, and write $G_{mn}(z, \theta\!=\!0)$ as
\begin{align}
G_{mn}(z) &= \frac{1}{4}\left[|U_{mp}(z)U_{nq}(z)+U_{mq}(z)U_{np}(z)|^2 \right.\nonumber \\
&\hspace{10mm} \left.+ I^p_m(z)I^p_n(z) + I^q_m(z)I^q_n(z) \right] , \label{seq5}
\end{align}
where $I_m^p(z)$ is the intensity at the $m$-th site for the initial excitation at the $p$-th site only.
Comparing  
Eqs.~\eqref{seq4} and \eqref{seq5}, we can write the relation between the intensity correlation and the photon number correlation as
\begin{align}
\Gamma_{mn}(z) = 4G_{mn} - I^p_m(z)I^p_n(z) - I^q_m(z)I^q_n(z). \label{seq6}
\end{align}
%%%%%%%%%%%%%%%%%%%%%%%%%%%%%%%%%%%%%%%%%%%%%
The probability $\mathfrak{p}_{m,n}$ of finding one particle at site $m$ and one at $n$ $(\neq m)$ 
at the output for the input coupling at site $p$ and $q \ne p$
is given by 
\begin{flalign}
&\mathfrak{p}_{m,n}(z) =|\bra{m,n}\hat{U}(z)\ket{p,q}|^2 && \nonumber\\
&=\Big|\dfrac{\bra{m}\otimes\bra{n} + \bra{n}\otimes\bra{m}}{\sqrt{2}}\hat{U}(z)\dfrac{\ket{p}\otimes\ket{q}+\ket{q}\otimes\ket{p}}{\sqrt{2}}\Big|^2 && \nonumber\\
&=|U_{mp}(z)U_{nq}(z) + U_{mq}(z)U_{np}(z)|^2 \nonumber\\
&= \Gamma_{mn}(z).  \label{seq7}
\end{flalign}
When we consider $n=m$, the probability is given by
\begin{flalign}
&\mathfrak{p}_{m,m}(z) =|\bra{m,m}\hat{U}(z)\ket{p,q}|^2 &&\nonumber\\
&=\Big|\bp{\bra{m}\otimes\bra{m}}\hat{U}(z)\bp{\dfrac{\ket{p}\otimes\ket{q}+\ket{q}\otimes\ket{p}}{\sqrt{2}}}\Big|^2 &&\nonumber\\
&=2|U_{mp}(z)U_{mq}(z)|^2 = \Gamma_{mm}(z)/2. &&  \label{seq8}
\end{flalign}
Hence, in the case of two identical particles in the SSH chain, the probability of finding one particle at $m$ and one at $n$ can be written in the following compact form 
\begin{align}
& |\tilde{\Psi}_{m,n}(z)|^2 = \Gamma_{mn}(z)/(1+\delta_{m,n}) \nonumber \\
&= \dfrac{1}{(1+\delta_{m,n})}\bs{4G_{mn} - I^p_m(z)I^p_n(z) - I^q_m(z)I^q_n(z)}, \label{seq9}
\end{align}
For the sake of presentation, $|\tilde{\Psi}_{m,n}(z)|^2$ is shown in a square grid in the main text. The modulus squared of the two-body wave function is then given by
\begin{align}
&|\Psi_{m,n}(z)|^2 = \Gamma_{mn}(z)/2 \nonumber \\
&= \dfrac{1}{2}\bs{4G_{mn} - I^p_m(z)I^p_n(z) - I^q_m(z)I^q_n(z)}. \label{seq10n}
\end{align}
with $|{\Psi}_{m,n}(z)|^2\!=\!|{\Psi}_{n,m}(z)|^2$. In this case, both indices $(m$ and $n)$ run from $1$ to $N$, where $N$ is the total number of sites in the lattice.

{\it {The fermionic case $-$}} In the case of two identical fermions, it is straightforward to show that the quantum correlation in Eq.~\eqref{seq4} will be modified as
\begin{align}
\Gamma^f_{mn}(z) = |U_{mp}(z)U_{nq}(z)-U_{mq}(z)U_{np}(z)|^2 . \label{seq12}
\end{align}
Here, the superscript $f$ indicates that we are considering the two-body fermionic states. Notice that the negative sign in Eq.~\eqref{seq12} arises due to fermionic statistics, which leads to destructive interference under particle exchange. 
To experimentally simulate the above two-fermion correlation, we consider $\theta\!=\!\pi$ in Eq.~\eqref{seq5-0} and write $G_{mn}(z, \theta\!=\!\pi)$ as
\begin{align}
G^f_{m,n}(z) = \dfrac{1}{4}\big[\Gamma^f_{m,n}(z)+I^p_m(z)I^p_n(z)+I^q_m(z)I^q_n(z)\big]
\label{GmnF2}
\end{align}
Eq.~\eqref{GmnF2} mathematically connects the quantum correlation $\Gamma^f_{m,n}(z)$ and the intensity correlation $G^f_{m,n}(z)$. The modulus squared of the two-fermion wave function is then given by
\begin{align} \label{ImnF3}
|\Psi^f_{m,n}(z)|^2 = \dfrac{1}{2}\big[4G^f_{m,n}(z) - I^p_m(z)I^p_n(z)-I^q_m(z)I^q_n(z)\big]
\end{align}
We can thus experimentally emulate the modulus squared of
the two-fermion wave function by combining the measurements of intensities and intensity
correlations.

In summary, we obtained the mathematical connection between the quantum (bosonic as well as fermionic) %{\color{red}$\Gamma_{m, n}$} 
and intensity correlations. %{\color{red}$G_{m, n}$ from Eq.~\ref{seq4} and Eq.~\ref{seq5}}
The purpose of experimentally constructing the matrix $G_{m, n}$ ($G^f_{m, n}$) is to obtain the two-body observable $|\Psi_{m,n}|^2$ ($|\Psi^f_{m,n}|^2$), %using Eq.~(3) in the main text, 
which is otherwise not possible by looking at the intensities $I_m$ for different phase realizations.

%%%%%%%%%%%%%%%%%%%%%%%%%%%
% End Derivation of Eq 3

%\section{Fabrication and characterization details}

%%%%%%%% Experimental details %%%%%%%
\begin{figure}[ht]
    \centering
    \includegraphics[width=1\linewidth]{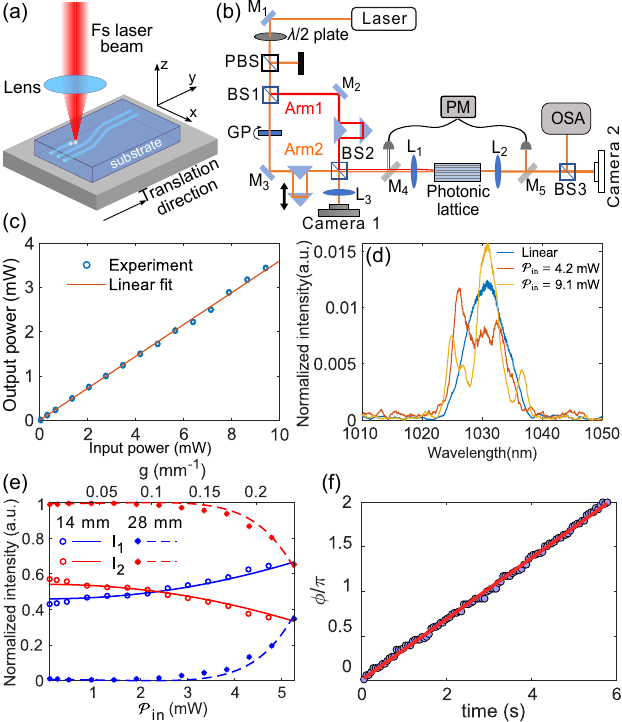}
    \caption{(a) A simplified cartoon of femtosecond laser writing. (b) Experimental set-up for preparing the two-site input state with a tunable relative phase. Here, $M_{1-5}$ are silver-coated mirrors ($M_{4-5}$ are mounted on flip-mounts), BS1-3 are beam splitters, $L_{1-3}$ are convex lenses, PM is a power meter, OSA is an optical spectrum analyzer. The optical delay lines are used to temporally synchronize the laser pulses of arm $1$ and arm $2$. The relative phase is tuned by rotating a glass plate (GP) mounted on a motorized rotational stage.
    (c) The variation of average output power as a function of average input power for a typical nonlinear experiment. The linear behavior implies negligible nonlinear loss due to multi-photon absorption processes.
    (d) Wavelength spectra in the linear and nonlinear regimes.
    (e) Nonlinear characterization of two straight coupled waveguides with centre-to-centre spacing of $29.5$ $\mu$m. Two sets of data correspond to two interaction lengths, $z_{\text{max}}\!=\!14$ mm and $28$ mm. The intensity at the waveguide where the light was launched at $z\!=\!0$ is denoted by $I_1$. (f) {Measured variation of the relative phase $\phi$ as the glass plate shown in 
    % Supplementary Fig.~\ref{fig_setup}
    (b) is rotated with a fixed angular velocity. The solid line is a linear fit. The standard deviation of the phase is $\Delta \phi\!=\!\pi/30$.} 
    }
    \label{fig_setup}
\end{figure}

\begin{figure*}[ht]
\includegraphics[width=0.85\textwidth]{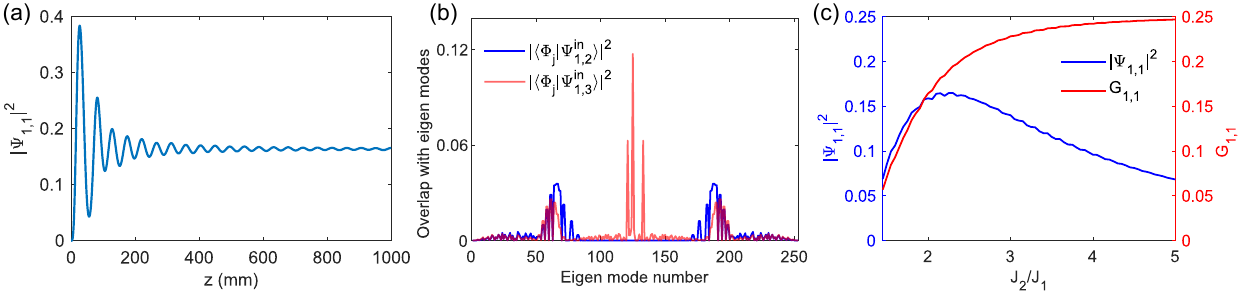} 
\caption{(a) The variation of $|\Psi_{1,1}(z)|^2$ as a function of $z$ for the experimentally realized values of $J_1$ and $J_2$.
(b) The overlap of the input state $\Psi^{\text{in}}_{1,3}$ and $\Psi^{\text{in}}_{1,2}$ with the two-body eigenstates $\phi_j$. Here, $J_{1,2}\!=\!\{0.042, 0.096\}$ mm$^{-1}$. (c) The variation of $|\Psi_{1,1}|^2$ and $G_{1,1}$ at $z\!=\!1000\, {\text{mm}}$ as a function of $J_{2}/J_1$.
  }
\label{BIC_sup}
\end{figure*}

\section{Fabrication and characterization details}\label{fab_char}

In this section, we briefly discuss how the photonic lattices were fabricated and characterized.

{\it Fabrication details:}
Waveguide-based photonic devices were fabricated using femtosecond laser writing%{\color{red}[56, 57]}
~\cite{davis1996writing, szameit2010discrete} 
-- a laser-based technique to modify refractive index inside a transparent and dielectric medium. Optical pulse trains of $260$~fs FWHM at $1030$ nm central wavelength and $500$~kHz repetition rate were generated from a commercially available Yb-doped fiber laser (Satsuma, Amplitude Laser Inc.). The power and polarization of the laser beam were controlled by using a polarizing beam splitter and wave plates. For the fabrication purpose, the laser beam with circular polarization was focused using a $0.4$ NA lens %(A110TM-B, Thorlabs Inc.) 
inside a borosilicate (BK7) glass substrate mounted on high-precision $x$-$y$-$z$ translation stages (Aerotech Inc.); see 
% Supplementary
Fig.~\ref{fig_setup}(a). 
% S1(a) 
To create each waveguide, the substrate was translated once through the focus of the laser beam at $6$~mm/s speed and $185$ mW average power. 
%
%The translation speed of fabrication and laser power were optimized to create low-loss single-mode waveguides at a depth of $100 \, \mu$m from the top surface of the substrate. 
%
The translation speed of fabrication and laser power were optimized to create low-loss single-mode waveguides 
supporting a nearly circularly symmetric and tightly-confined fundamental mode
at a depth of $100 \, \mu$m from the top surface of the substrate. 
%{\color{magenta}Fabrication parameters were optimized to obtain low-loss single mode waveguides supporting a nearly circularly symmetric and tightly-confined fundamental mode.}
The mode field diameters ($1/e^2$ of the intensity peak) of the fundamental mode were measured to be $25 \, \mu$m and $24.5 \, \mu$m along the vertical and horizontal directions, respectively, at $1030$ nm wavelength of light.
The propagation loss was estimated by subtracting the coupling losses from the measured insertion loss and was found to be $0.34$~dB/cm.
The coupling between two waveguides were controlled by varying the separation between them.
The off-diagonal disorder in the lattice was estimated by characterizing a set of two-waveguide devices with the same inter-waveguide spacing, and was found to be $\Delta J/J\!=\!6\%$. In addition, we observed a nearly full transfer of optical power in such two-waveguide devices, implying that the consisting waveguides have the same onsite energies. In other words, the diagonal disorder is negligible in our experiments.

{\it Characterization:} A schematic of the state preparation and device characterization is shown in 
% Supplementary
Fig.~\ref{fig_setup}(b). 
% Fig.~S1(b).
By using a grating pair (not shown in the schematic), near-transform limited $260$~fs laser pulses at $5$~kHz repetition rate were down-chirped and temporally stretched to $ 1.1$ ps. 
The polarizing beam splitter (PBS) transmits only the horizontal polarization of light whose power was tuned by rotating a $\lambda/2$ plate. 
The beam is split into two arms using a 50:50 beam splitter BS1. Optical delay lines are used in both arms to fine-tune the temporal synchronization of the pulses. A $1$-mm-thick glass plate (GP) mounted on a motorized rotational stage is placed in arm 2 to control the relative phase $\phi$ of the optical pulses. Specifically, we rotate the glass plate with a fixed angular velocity to vary the phase $\phi$ linearly in time in steps of $\pi/100$ with a  standard deviation of $\Delta\phi \!=\!\pi/30$, as shown in 
% Supplementary 
Fig.~\ref{fig_setup}(f). 
%Fig.~%\ref{supp_mat_phase}. 
% S2. 
The output intensity patterns were recorded at a frame rate of $18$ fps with an exposure time of $30$ ms. We also note that the power RMS stability of our laser source is $<\!1\%$, meaning the fluctuation in average power can be neglected.

Laser beams from arms 1 and 2 are reflected and transmitted, respectively, using another 50:50 beam splitter (BS2). The two beams are then coupled into two desired waveguides using a convex lens (L$_1$). 
The interference pattern produced by the two input beams is observed using camera~1 (CS165MU, Thorlabs Inc.), and the intensity distribution at the output of the photonic lattice is imaged on camera~2. 
During all measurements, the input and output powers were monitored by reflecting the light by the silver-coated mirrors $M_4$ and $M_5$ mounted on flip-mounts. The linear variation of the average output power with the average input power implies negligible nonlinear loss due to multi-photon absorption processes; see 
% Supplementary
Fig.~\ref{fig_setup}(c). 
% S1(c).
A small fraction of the output power is coupled to an optical spectrum analyzer (AQ6370D, Yokogawa) using the beam splitter BS3 for measuring the wavelength spectrum. In the nonlinear regime, the generation of new wavelength due to self-phase modulation (SPM) is shown in 
% Supplementary
Fig.~\ref{fig_setup}(d). 
% S1(d).
Because of temporal pulse shaping, the effect of SPM is minimized -- notice that the wavelength span in the nonlinear case is below $15$ nm, meaning the wavelength dependence of couplings can be ignored for the maximal power used in our experiments.
It should also be noted that the chromatic dispersion is negligible in our experiments because the maximal propagation distance is $80$ mm (the dispersion length for BK7 glass is $> 50$ m).

\textit{Estimation of nonlinear strength:} 
As mentioned in the main text, the strength of nonlinearity $g$ is tuned experimentally by varying the laser pulse energy, or equivalently, the average input power $\mathcal{P}_{\text{in}}$. 
To get a relation between $g$ and $\mathcal{P}_{\text{in}}$, 
we characterize nonlinear two-waveguide devices where the dynamics of the optical field is governed by
\begin{eqnarray}
\label{NL_coupler}
\hspace{-5mm}i\frac{\partial}{\partial z} \psi_{1(2)}(z)=\!-i\alpha \psi_{1(2)} -\!J \psi_{2(1)} -\!|\psi_{1(2)}|^2\psi_{1(2)} \; , \label{a1}
\end{eqnarray} 
where $\psi_{1(2)}$ is proportional to the slowly-varying complex amplitude of the optical field at the waveguide 1(2). Here, $\alpha$ is a measure of propagation loss, and the evanescent coupling $J$ was measured to be $0.059$ mm$^{-1}$. We considered two devices with the same inter-waveguide spacing $29.5 \, \mu$m and different interaction lengths $z_{\text{max}}\!=\!14$ mm and $28$ mm. For all measurements, the light was launched into waveguide 1, i.e., $|\psi_{1}(z\!=\!0)|^2\!=\!1$, and the variation of normalized intensities at the output of two waveguides are shown in 
% Supplementary
Fig.~\ref{fig_setup}(e).
For the known value of $J$ and $\alpha\!=\!0.0039$ mm$^{-1}$, we solved Eq.~\eqref{NL_coupler} and fitted the experimentally obtained data.
The top x-axis in 
% Supplementary
Fig.~\ref{fig_setup}(e) 
indicates the equivalent nonlinear strength, that is used for numerical fitting -- the solid and dashed lines. 
The nonlinear strength $g$ was found to be linearly proportional to the average input power $g\!=\!0.045  \, \mathcal{P}_{\text{in}}$  mm$^{-1}$mW$^{-1}$.

\section{More details on BICs}\label{bic_robust}
 
In the experiments, the BICs were probed by preparing an initial state, {$\Psi_{1,3}^{\text{in}}\!=\!\ket{1,3}$} and then observing a large value of $|\Psi_{1,1}|^2$ at the output. 
We note that the magnitude of $|\Psi_{1,1}|^2$ at a given $J_2/J_1>1$ varies as a function of the propagation distance.
In fact, $|\Psi_{1,1}|^2$ initially oscillates as a function of $z$, and then slowly approaches a nonzero value that is determined by the ratio $J_2/J_1$. The variation of $|\Psi_{1,1}(z)|^2$  for the experimentally realized values of $J_1$ and $J_2$ is shown in 
% Supplementary
Fig.~\ref{BIC_sup}(a).

Other than the BICs, the initial states $\Psi_{1,3}^{\text{in}}$ 
has a noticeable overlap with 
the semi-localized states associated with one particle on the edge and one on the bulk; see 
% Supplementary
Fig.~\ref{BIC_sup}(b) and 
% Figs.~2(a, b).
Figs.~\ref{fig_2}(a-d). 
However, as mentioned in the main text, the observed large value of $|\Psi_{1,1}|^2$ proves the existence of the BICs. To demonstrate that the large value of $|\Psi_{1,1}|^2$ can not arise due to the excitation of the semi-localized states, we consider another initial state {$\Psi_{1,2}^{\text{in}}=\ket{1,2}$} that has a large overlap ($80\%$) with the semi-localized states; however, it does not excite the BICs; see 
% Supplementary
Fig.~\ref{BIC_sup}(b). As a result, non-zero elements $|\Psi_{1,n}|^2$ and $|\Psi_{m,1}|^2$ ($m, n \ne 1$) are observed in the experiments with nearly zero magnitude of $|\Psi_{1,1}|^2$; see 
% Figs.~3(d, e).
Figs.~\ref{fig_2b}(d, e).

The spatial extent of the BICs is determined by the ratio of $J_2$ and $J_1$ -- as $J_2/J_1$ increases, the BICs become more localized. However, the amplitude of the BICs at $\{1,3\}$ and their overlap with the input state $\Psi_{1,3}^{\text{in}}$ is maximal near $J_2/J_1\!=\!2.3$. As a result, $|\Psi_{1,1}(z\!=\!1000 \,{\text{mm}})|^2 $ exhibits a peak around $J_2/J_1\!=\!2.3$, as shown in 
% Supplementary
Fig.~\ref{BIC_sup}(c).
 To obtain a good signature of the BICs in our experiments, %$J_{1,2}\!=\!\{0.042, 0.096\}$ mm$^{-1}$ 
$J_{1,2}$ were chosen 
to be $\{0.042, 0.096\}$ mm$^{-1}$ -- these couplings have been used in
all calculations unless otherwise mentioned.
It should also be noted that unlike $|\Psi_{1,1}|^2$, the intensity correlation at \{1,1\}, $G_{1,1}$ increases with $J_2/J_1$ and then converges to $0.25$.

\begin{figure}[ht]
\includegraphics[width=1\linewidth]{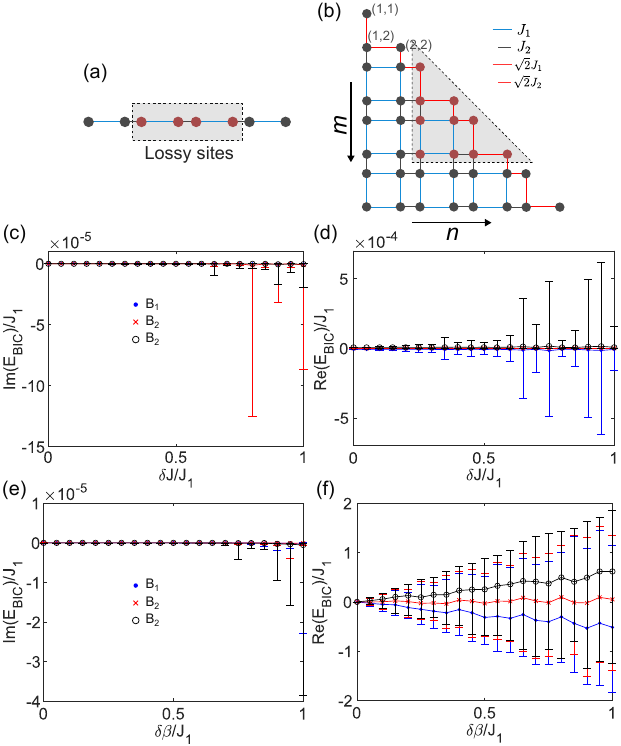}
\caption{(a) The SSH chain with a total $N\!=\!N_s+N_e$ sites where $N_e$ is the number of lossy sites in the middle. $N_s/2$ sites on both edges have zero onsite loss. (b) Two indistinguishable bosons in (a) can be mapped to a single particle in the 2D lattice.
(c, d) The variation of the imaginary and real part of the energy of the BICs as a function of coupling disorder. The data points and the error bars indicate the mean and maximum deviation of $100$ realizations of disorder. (e, f) Same as (c, d) for the onsite disorder. Here, $J_{1,2}\!=\!\{0.042, 0.096\}$ mm$^{-1}$. 
}
\label{BIC_sup_disorder}
\end{figure}

{\it Robustness of BICs.$-$} To inspect the localization feature of the BICs shown in 
% Figs.~2(b),
Figs.~\ref{fig_2}(b-d), 
we introduce onsite loss in the middle of the SSH array, 
% Supplementary
Figs.~\ref{BIC_sup_disorder}(a, b) {\color{TriClr} -- the two-body Hamiltonian in Fig.~\ref{BIC_sup_disorder}(a) is equivalent to a single-particle Hamiltonian in a triangular-shaped lattice shown in Figs.~\ref{BIC_sup_disorder}(b).} To be more specific, we consider total $N\!=\!N_e+N_s\!=\!32$ sites in the array, where $N_s/2\!=\!10$ sites on both edges are loss-less and $N_e\!=\!12$ sites in the middle have onsite loss $\gamma\!=\!J_1/10$. In this situation, all extended eigenstates of the system have a finite imaginary eigenvalue. 
For our system parameters, the imaginary parts of the BIC eigenvalues were found to be ${\text{Im}}(E_{\text{BIC}})/J_1\!\le\!10^{-8}$, which converges exponentially to zero with increasing system size $N_s$
% ~[35]. 
~\cite{cerjan2020observation}. 
This observation confirms that the BICs are exponentially localized to the edges.

Although BICs can appear in various experimental settings, they are usually fragile states, meaning small perturbations can destroy them. However, the two-body BICs we demonstrate in the main text show remarkable robustness against disorder.
To inspect the robustness of the BICs, we first consider off-diagonal disorder, i.e., disorder in couplings. Each coupling in the lattice is modified as $\tilde{J}_{1, 2}\! =\! J_{1, 2}+\Delta J$, where the random variation in couplings $\Delta J$ is taken from a uniform distribution in the interval $[-\delta J, \delta J]$.  As shown in Figs.~\ref{BIC_sup_disorder}(c, d), the mean value of real as well as imaginary eigenvalues of the BICs are not affected by coupling disorder $\lesssim J_1/2$.
On the other hand, when diagonal disorder (i.e., disorder in onsite energies, $\tilde{\beta} \! = \! \beta+\Delta \beta$) is introduced, the BICs become non-degenerate, however, they still remain localized and embedded in the two-body bulk band as long as the disorder strength $\lesssim J_1/2$, see Figs.~\ref{BIC_sup_disorder}(e, f). It should be noted that the width of the middle two-body bulk band is $\pm 2J_1$.

\begin{figure}[t]\includegraphics[width=1\linewidth]{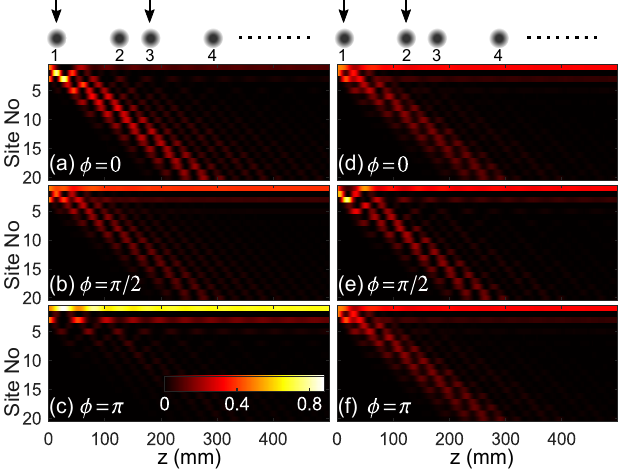}
    \caption{(a-c) The linear dynamics in the SSH lattice when light is coupled to waveguides 1 and 3 (as indicated in the sketch on the top) for 3 different relative phases indicated on each image. The overlap of the input state with the edge states of the SSH lattice is maximal (minimal) for $\phi\!=\!\pi$ $(0)$. (d-f) Same as (a-c) for input excitation at waveguides 1 and 2. In this case, the light localized in the edge waveguide does not change with $\phi$. In each image, the field of view is smaller than the actual numerics, where 50 waveguides were considered. At all propagation distance the total power is normalized to 1.}
    \label{fig_lin_dynamics}
\end{figure}

\begin{figure}[h!]
    \includegraphics[width=1\linewidth]{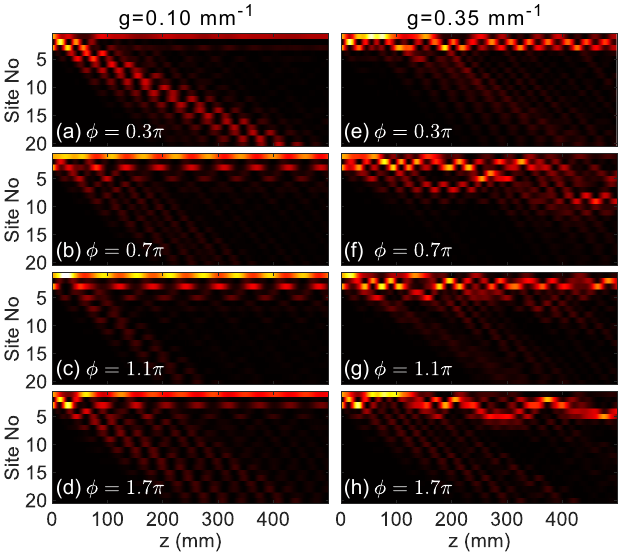}
    \caption{Numerically calculated dynamics of the two-site input state for four different relative phases, $\phi\!=\!0.3\pi, \, 0.7\pi, \, 1.1\pi $ and $1.7\pi$ indicated on each image. The nonlinear strength is $g\!=\!0.1$ mm$^{-1}$ for (a-d) and $g\!=\!0.35$ mm$^{-1}$ for (e-h). An edge soliton and long-lived breathers are observed in (a) and (b-d), respectively. The light tends not to be maximally localized on the edge site $1$ in (e-h). The filed of view is smaller than the actual numerics where $50$ waveguides were considered. At all propagation distance the total power is normalized to $1$. The color-map is same as in Fig.~\ref{fig_lin_dynamics}.}
    \label{fig_NL_dynamics}
\end{figure}

\section{Propagation calculations}

In this section, we briefly discuss how the propagation of light is numerically calculated.
We consider a $500$-mm-long SSH lattice of 50 waveguides, which has the same couplings as found experimentally.
The linear dynamics is shown in Figs.~\ref{fig_lin_dynamics}(a-c) [see also Supplementary Movie 1] for input excitation at waveguides 1, and 3 with relative phase $\phi = 0, \pi/2$ and $\pi$, respectively -- only the first 20 sites are shown here. 
By varying the relative phase, the overlap of the input state with the edge states can be tuned in this case.
For example, in the case of  $\phi \!=\! 0$, a small amount of light remains on the edge   because of the small ($13\%$) edge overlap.
On the other hand, the edge overlap is large ($84\%$) for $\phi \!=\! \pi$, and hence, the light remains localized mostly on the edge.
In the experiments presented in 
% Fig.~3(a),
Fig.~\ref{fig_2b}(a), 
we measured output intensities after a propagation of $80$ mm, which is in good agreement with the numerical results in 
% Supplementary
Figs.~\ref{fig_lin_dynamics}(a-c).

The linear dynamics for input excitation at waveguides 1, and 2 with three different relative phases are shown in 
% Supplementary
Figs.~\ref{fig_lin_dynamics}(d-f). Irrespective of the phase value, the same amount of light is localized on the edge waveguide.
This can be understood from the fact that the edge states of the SSH lattice live on the odd sites ($1, 3, 5, \dots$), as shown in 
inset of Fig.~1(c).
% {\color{red}Fig.~\ref{fig_1}(h)}.
In other words, the overlap between the input state and the edge states does not change as the relative phase is varied.

%%% figure
\begin{figure}[]
    \includegraphics[width=0.7\linewidth]{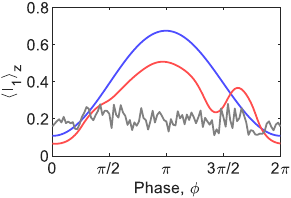}
    \caption{The $z$ averaged intensity of light in waveguide 1 for the linear, weak and strong nonlinear regimes shown by blue, red and grey curves, respectively. Here $z_{\text{max}}\!=\!500$~mm.}
    \label{integration_of_I1}
\end{figure}

\begin{figure*}[t!]
\includegraphics[width=0.78\textwidth]{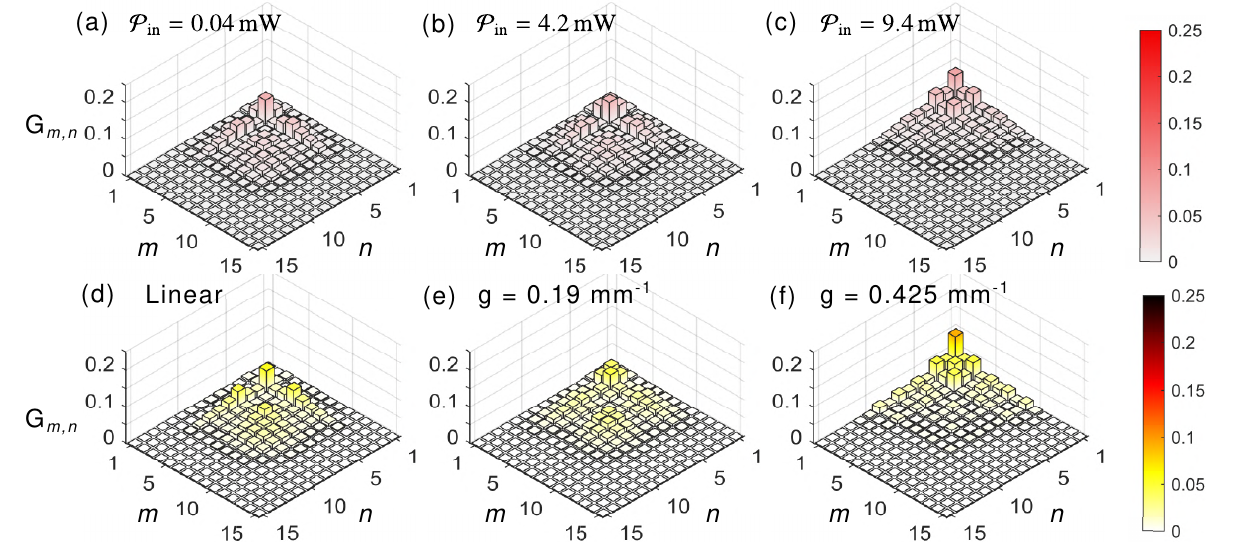}
\caption{Experimentally measured (a-c) and numerically calculated (d-f) variation of ${G}_{m,n}$ as a function of nonlinearity for a  lattice with homogeneous coupling strength $J\!=\!0.046$ mm$^{-1}$. Here, $z_{\text{max}}$ is $80$ mm, and the loss coefficient is $\alpha\!=\!0.0039$ mm$^{-1}$.
}
\label{int_corre}
\end{figure*}

We now consider the nonlinear dynamics of the two-site input state launched at waveguides 1, and 3 for  $g\!=\!0.1$ mm$^{-1}$ and $g\!=\!0.35$ mm$^{-1}$, see 
% Supplementary
Fig.~\ref{fig_NL_dynamics} and Supplementary Movie 2 and 3. 
As discussed in the main text,
at a weak nonlinear strength $g\!=\!0.1$ mm$^{-1}$, long-lived breathers and solitons are formed at the edge of the lattice. %for a wide range of relative phase $\phi$. 
% Supplementary
Figs.~\ref{fig_NL_dynamics}(a) and (b-d) confirm the formation of edge soliton and breather-like solutions, respectively.
In case of 
% Supplementary
Fig.~\ref{fig_NL_dynamics}(a), most of the light ($\approx 75 \%$) penetrates the bulk because the edge soliton requires relatively low power. {In our propagation calculations, such long-lived breathers were clearly observed for a nonlinear strength ranging from $g\!=\!0.05$ mm$^{-1}$ to $0.12$ mm$^{-1}$.}

%As shown in Supplementary Fig.~\ref{breather_fig}, edge solitons break down as the optical power exceeds a threshold value.  
%

For a larger nonlinearity $g\!=\!0.35$ mm$^{-1}$, the light tends not to be maximally localized on the edge site $1$ as shown in 
% Supplementary
Figs.~\ref{fig_NL_dynamics}(e-h). Unlike the weak nonlinear regime, in this case, the nonlinear dynamics is extremely sensitive to the relative phase $\phi$. It should be highlighted that true breathers can form in the large nonlinear regime as shown in 
% Supplementary
Fig.~\ref{breather_fig}.

In the main text, we have presented the variation of intensity at waveguide 1 as a function of phase $\phi$. The variations of $I_1(\phi)$ clearly show distinct behavior depending on the strength of the nonlinearity, 
% Fig.~4(e).
Fig.~\ref{fig_3}(e). 
To show that this behavior is not a $z$ dependent effect, we consider the $z$ averaged intensity of light in  waveguide 1, i.e. $\langle I_1\rangle_z  \!=\!\frac{1}{z}\int_0^z \text{d}z'I_1(z')$. In 
% Supplementary
Fig.~\ref{integration_of_I1}, we show $\langle I_1\rangle_z$ for three different nonlinear strengths. Notice the appearance of a single peak and a double peak in the linear and weak nonlinear ($g\!=\!0.1$~mm$^{-1}$) regimes respectively. For a large nonlinearity, $\langle I_1\rangle_z$ fluctuates rapidly with phase, as shown by the grey curve in 
% Supplementary
Fig.~\ref{integration_of_I1}.

\section{Intensity correlations in a homogeneous 1D array}

In Fig.~\ref{int_corre}, we show intensity correlations $G_{m,n}$ for a lattice with homogeneous coupling, $J = 0.046$ mm$^{-1}$. Figs.~\ref{int_corre}(a-c) show the experimentally obtained correlations, and Figs.~\ref{int_corre}(d-f) show the corresponding numerical results that are in good agreement with experiments. 
For this array, there is no state localized on the edge, and hence, $G_{m, n}$ exhibits delocalization in the linear case. As the strength of nonlinearity is increased, the intensity correlations become more localized near the edge.
Note that we observed an opposite behavior, i.e., a localization-to-delocalization feature with increasing nonlinearity, in case of SSH lattice (
 % Fig.~5
 Fig.~\ref{fig_3}
 in the main text).
In numerics, we introduced loss by adding a term $-i\alpha \psi_j$ on the right-hand side of Eq.~\ref{a1},
%\ref{SE}, 
where $\alpha$ is the loss coefficient that was experimentally measured to be  $\alpha\!=\!0.0039$ mm$^{-1}$.

\section{Discrete breathers}\label{DB}

As mentioned in the main text, a discrete breather is a nonlinear state (usually exponentially) localized along the lattice axis and periodic along the propagation distance. Here, we are considering a photonic SSH lattice described by a static, i.e., $z$-independent linear Hamiltonian $\hat{H}^{\text{lin}}$. When a breather is formed, the nonlinearity locally modifies the onsite energy in a $z$-periodic manner. In other words, the total Hamiltonian $\hat{H}_t\!=\!\hat{H}^{\text{lin}}+\hat{H}^{\text{nl}}$ is periodic in $z$ with a periodicity of the breather, i.e., $\hat{H}_t(z+z_0)=\hat{H}_t(z)$, where the period $z_0\!=\!2\pi/\Omega$.
It is evident that the Floquet theory can be employed to obtain the breather solution and its quasienergy.

\begin{figure}[t!]
\includegraphics[width=1\linewidth]{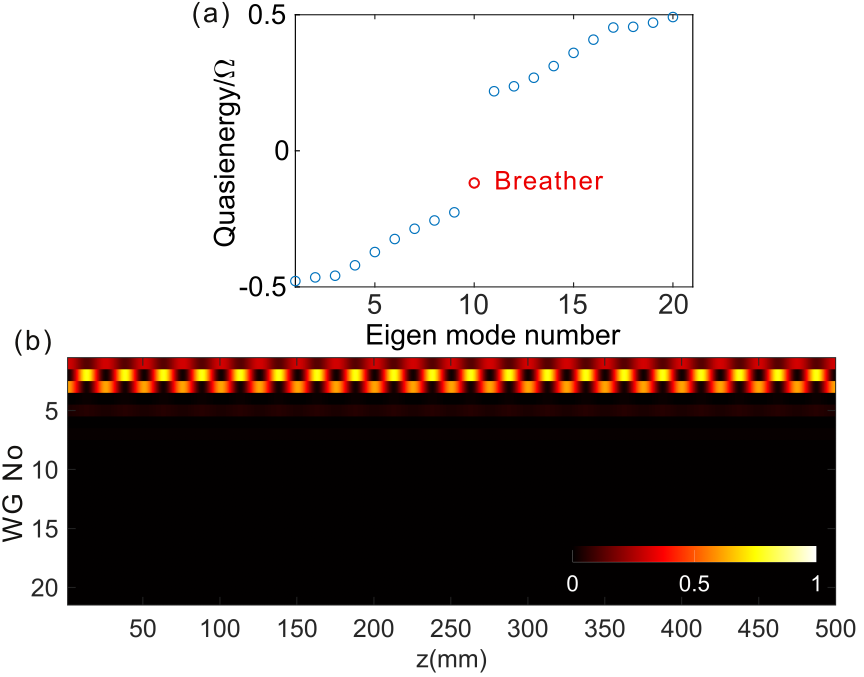}
\caption{(a) Quasienergy spectrum of the SSH lattice supporting a discrete breather. The linear modes are indicated in blue, and the breather quasienergy is shown in red. (b) Propagation of the breather in (a). Here, the period of the breather is $25$ mm, and the nonlinear strength is $g\!=\!0.31$~mm$^{-1}$}
\label{breather_fig}
\end{figure}

The propagation of optical intensity presented in Supplementary Movie 3 shows breather-like edge modes for a few values of the relative phase $\phi$. Inspired by this observation, we use a Floquet self-consistency algorithm
% ~[68]
~\cite{lumer2013self} 
to look for true breathers. We consider a finite SSH lattice of $21$ sites and experimentally realized values of couplings. Starting from a `guess' solution, we iteratively calculate the breather state, bifurcating from the linear topological edge mode. The quasienergy of the breather and its propagation dynamics is shown in Figs.~\ref{breather_fig}(a, b). The breather is localized on the edge with a periodicity of $z_0\!=\!25$~mm.

Let us now consider the breather-like solutions shown in 
% Fig.~4(f)
Fig.~\ref{fig_3}(f) 
and Supplementary Movie 2. We note that these solutions are observed for a wide range of relative phases in the propagation calculation. Importantly, the period of such states is observed to be $\approx 58$ mm, i.e., 
more than twice the true breather shown in Fig.~\ref{breather_fig}(b). Now, the periodic nature of the Hamiltonian makes the quasienergy spectrum periodic, i.e., $\varepsilon \in [-\Omega/2, \Omega/2]$. For a long period of $58$ mm, the quasienergy spectrum becomes gapless due to band folding. Hence, the  breather-like states with long periods of oscillations are in resonance with the linear Floquet modes, causing their finite lifetime. In other words, these breather-like states slowly radiate power into the bulk during propagation.

\section{Description of supplementary movies}

{\it Supplementary Movie 1.$-$} 
The linear ($g\! \rightarrow \! 0$) propagation of optical intensity in the SSH lattice for the two-site initial excitation at waveguides 1 and 3 with relative phase $\phi$ indicated on the top. The field of view is smaller than the actual numerics, where $50$ waveguides were considered. Here, we have considered experimentally realized values of $J_{1, 2}$.

{\it Supplementary Movie 2.$-$} Nonlinear propagation at $g= 0.1$ mm$^{-1}$, otherwise same as Supplementary Movie 1. 
Notice the formation of long-lived discrete breathers near 
$\phi=(0.9 \pm 0.4)\pi$ and $\phi=(1.65 \pm 0.1)\pi$.

{\it Supplementary Movie 3.$-$} Nonlinear propagation at $g= 0.35$ mm$^{-1}$, otherwise same as Supplementary Movie 1. Notice that the dynamics is more sensitive to relative phase $\phi$, and  localized nonlinear edge modes are formed only for a few specific values (not for a wide range) of the relative phase $\phi$.

%\bibliography{ref-NL-corr-lpr}


\end{document}